\documentstyle[aps,epsf,twocolumn]{revtex}
\input epsf.sty
\begin{document}
\draft
\flushbottom
\twocolumn[
\hsize\textwidth\columnwidth\hsize\csname @twocolumnfalse\endcsname

\title{Superconductivity, Josephson coupling and order parameter symmetry in
striped cuprates}
\author{A.~H.~Castro~Neto $^1$ and F.~Guinea $^2$}

\address{$^1$ Department of Physics,
University of California,
Riverside, CA, 92521 \\
$^2$Instituto de Ciencia de Materiales, Consejo Superior de Investigaciones
Cient{\'\i}ficas,
Cantoblanco, E-28049, Madrid, Spain}
\date{\today}
\maketitle
\tightenlines
\widetext
\advance\leftskip by 57pt
\advance\rightskip by 57pt

\begin{abstract}
We consider a renormalization group study of the problem of coupled stripes of
holes in cuprates. We use a
model of a mesh of horizontal and vertical stripes and study
the problem
of superconductivity via the Josephson coupling.
We discuss the evolution of the Fermi surface with doping and
temperature, the existence
of Luttinger and/or Fermi liquid behavior, the presence of pre-formed
Cooper pairs and the symmetry and magnitude of the superconducting order
parameter.
\end{abstract}
\pacs{PACS numbers:74.20.Mn, 74.50.+r, 74.72.Dn, 74.80.Bj}

]
\narrowtext
\tightenlines

After more than ten or so years after their discovery,
there is still a debate about the microscopic physics of
high temperature superconductors. Due to
strong electronic correlations transition metal oxides
show a rich variety of phenomena such as
anomalous transport, incommensurate magnetic fluctuations
and phase separation.
Incommensurate magnetic effects have been observed in
La$_{2-x}$Sr$_x$CuO$_{4}$ \cite{LSCO} and have been shown to have their origin
in
the formation of striped phases in related compounds such as
La$_{1.6-x}$Nd$_{0.4}$Sr$_x$CuO$_{4}$
\cite{tranquada} and antiferromagnetic insulators such as
La$_{2-x}$Sr$_x$NiO$_{4+y}$ \cite{nickel}.
Recently it has been reported that
the same incommensurate effects are observed in YBa$_2$Cu$_3$O$_{7-\delta}$
and Bi$_2$Sr$_2$CaCu$_2$O$_{8-x}$ \cite{YBCO}.
Furthermore, various different experiments point
towards a {\it low dimensional} picture of these compounds and to a closeness
to quantum critical behavior which can be interpreted in terms of static
stripe formation \cite{aeppli}.
In the stripe picture
the low dimensionality is due to charge segregation
into domain walls \cite{steve}. Incommensurability appears
because of the magnetic domain formation.
In the antiferromagnetic phase stripes seem to explain well the available data
\cite{anti}. In this paper we study the problem of superconductivity
and Fermi surface evolution with doping and temperature in
a model proposed recently in which the stripes
form a mesh of horizontal and vertical Luttinger liquids that hybridize with
each other \cite{eu}.
A similar model was proposed sometime ago in order
to study the crossover from repulsively interacting
one-dimensional Luttinger liquids to two-dimensional Fermi liquids \cite{paco}
and has similarities with models of c-axis tunneling \cite{anderson}.

Our analysis is based on renormalization group
calculations \cite{shankar} and we show that the
model not only provides a very simple picture of the Fermi surface in
underdoped cuprates
but also agrees with recent experiments, especially angle resolved
photo-emission
(ARPES) \cite{shen,randeria,schrieffer}.
We also calculate the relevant energy scales for the model and propose
new experiments that can test our ideas. Our model differs substantially
from earlier work on the effects of stripes in ARPES \cite{salkola}.

\begin{figure}
\epsfysize7 cm
\hspace{0cm}
\epsfbox{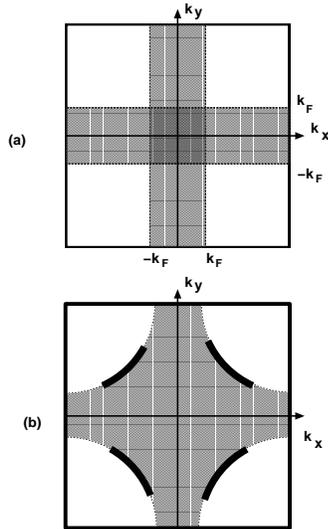}
\caption{(a) Fermi surface for the stripes without hybridization; (b)
Fermi surface with hybridization: the continuous line shows the region
of Fermi liquid behavior (finite residue) and the doted line shows region of
non-Fermi liquid
behavior.}
\label{fermi}
\end{figure}

The model consists of horizontal and vertical stripes in
different CuO$_2$ planes
described as Luttinger liquids with a characteristic
interaction parameter $g$ where $g>1$ ($g<1$) for attractive (repulsive)
interactions \cite{boso} and separated from each other by antiferromagnetic
regions. The Hamiltonian of the problem is written as (we use units such that
$\hbar=k_B=1$)
\begin{eqnarray}
{\cal H} = \sum {\cal H}_{x , n_y } + \sum {\cal H}_{y , n_x } +
{\cal H}_T
\nonumber
\end{eqnarray}
where ${\cal H}_{x n_y}$ describes a 1D chain with lattice spacing $a$
in the $x$ direction,
which intersects the $y$-axis at $n_y \ell$ where $n_y$ is an integer
and $\ell=N a$ is the inter-stripe distance.
A similar notation is used
to define ${\cal H}_{y , n_x}$. ${\cal H}_T$ describes tunneling between
vertical and horizontal chains with amplitude $t$.
We assume that the coupling is weak and treat
${\cal H}_T$ as a perturbation on the Luttinger liquid.
We can define an extended Brillouin zone, in terms
of the lattice constant of each individual chain, $a$.
Each set of chains gives rise to a band in the extended zone.
In the absence of hybridization the electrons are localized on the stripes
and the Fermi surface is shown in
Fig.\ref{fermi}(a) where the horizontal and vertical bands are
bounded by the lines at $\pm k_F$ where $k_F = \pi n/2$ is the Fermi momentum
for a chain with linear density $n$.

The scaling equations for the coupling constants are obtained by
tracing out high energy degrees and rescaling towards low energies
\cite{shankar}.
The scaling equation for the tunneling is given by
\begin{eqnarray}
\partial_{l} \tilde{t} = \tilde{c}_4 \tilde{t}
\nonumber
\end{eqnarray}
where $\tilde{c}_4=1-1/\tilde{g}$ where $1/\tilde{g}=(g+1/g)/4$
and $\tilde{t}$ is the ratio between the hopping in ${\cal H}_T$
and the infrared cutoff,
$\Lambda$ ($d l = d|\Lambda|/\Lambda$) \cite{kf}.
The renormalized tunneling energy is
$t_R = t (t/\omega_c)^{(1-\tilde{c}_4)/\tilde{c}_4} (a/\ell)^{1/\tilde{c}_4}$
where $\omega_c$ a bandwidth cut-off \cite{paco}.
In the absence of tunneling, the electrons have a well defined momentum
in the direction parallel to the stripe where they are located.
${\cal H}_T$ induces the folding of the Brillouin zone shown in
Fig.\ref{fermi}(a) into $N \times N$ smaller zones. However, these
are higher order effects in terms of the tunneling and distortions
of the mesh, induced either by static disorder or thermal fluctuations
\cite{zaanen}, will smear these additional ``shadow" bands \cite{shadow}.
Inter-chain tunneling will also hybridize the states derived from horizontal
and
vertical bands.
Observe that for a doping amount of $x$ holes there are
$1/N-x$ electrons participating on this Fermi surface. The other
$1-1/N$ electrons form the surrouding antiferromagnetic background.
Since the antiferromagnetic and stripe electrons are in equilibrium
they must have the same chemical potential (so that the lower
Hubbard band of the antiferromagnet is pinned at the stripe Fermi
surface). Thus, superimposed to the Fermi surface of Fig.\ref{fermi}(a), one
must have a diamond shaped, and somewhat blurred,
occupation in momentum space for the antiferromagnetic
electrons, with area proportional to $1-1/N$. The final result is
shown on Fig.\ref{fermi}(b).
The total area seen by ARPES
should be proportional to $1-x$, corresponding to a large Fermi surface.
These results agree well with the available ARPES data
\cite{shen,randeria}.

In most points of the Brillouin zone, the horizontal and
the vertical bands are not degenerate and there is
a splitting $\delta_{k_x,k_y}$ between bands.
At zero temperature $\delta_{k_x,k_y}$ cuts off the RG flow.
If $\delta_{k_x,k_y} > t_R$ the vertical and horizontal
stripes decouple and coherence effects are not
expected.
As shown in ref.\cite{paco} the quasiparticle residue vanishes in the
regions close to $(\pm \pi/a,0)$ and $(0,\pm \pi/a)$ and
the system behaves like a {\it Luttinger liquid}.
The only exception are the
points close $(\pm k_F ,\pm k_F)$ where $\delta_{\pm k_F,\pm k_F}=0$.
When $\delta_{k_x,k_y} < t_R$ the horizontal and vertical bands will
become strongly hybridized,
and a quasiparticle residue is finite \cite{Wen} leading to local
{\it Fermi liquid} behavior.
The {\it crossover} from Luttinger to Fermi liquid along the Fermi surface is
defined by $t_R \approx \delta_{k_x,k_y}$ as shown in Fig.\ref{fermi}(b).
It turns out however
that the regions of non-Fermi liquid behavior are nested and therefore will be
subjected to instabilities \cite{pines}.
An interesting consequence of this scenario is that
there is a {\it crossover} from Luttinger liquid to Fermi
liquid behavior also as a function
of $\ell$ as $\ell$ decreases. There is a minimum value of
$\ell$, say $\ell^*$, for which
$\delta(\ell^*) \approx t_R(\ell^*)$. Since $\ell$ is a monontonic
decreasing function of $x$ \cite{eu} there is
a crossover doping $x^*$ such that for $x<x^*$ ($x>x^*$) we have Luttinger
(Fermi) liquid behavior.

\begin{figure}
\epsfysize5 cm
\hspace{0cm}
\epsfbox{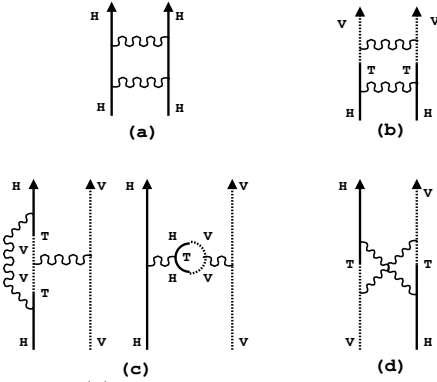}
\caption{(a) Cooper pair diagram within a horizontal stripe (H); (b)
Cooper pair diagram for a pair in a horizontal stripe (H) tunneling (T) to
a vertical stripe (V); (c) direct processes; (d) exchange processes.}
\label{diagram}
\end{figure}

Interaction processes do not need to involve an energy cost, even if they
conserve crystal momentum. We can use a standard diagrammatic analysis
to study
the interactions which can be generated by the combined effect
of the intra-chain couplings and the inter-chain hopping.
The irreducible diagrams that can be built starting with two electrons in
either the horizontal or the vertical chains, which interact
between them, hop into the other set of chains, and interact again.
Besides the usual pairing process within each chain which is shown in
Fig.\ref{diagram}(a)
there are two types of possible couplings which satisfy momentum
conservation which are depicted in Fig.\ref{diagram}(b) :
{\it i)} The Cooper pair channel: Cooper pair in the horizontal
chains is transferred to the vertical chains, or vice-versa;
{\it ii)} The transfer of  Cooper pairs with total
momentum $( \pm 2 k_F , \pm 2 k_F )$.  There are also
{\it direct} (Fig.\ref{diagram}(c))
and {\it exchange} (Fig.\ref{diagram}(d)) diagrams which give a vanishing
contribution ibecause of phase space restrictions\cite{shankar}.
These processes can all be written as
$\hat{A}_{vh} = \hat{O}^\dag_v \hat{O}_h$, where $\hat{O}$
is an operator which creates, or destroys, a pair of electrons in one chain.
As each separated chain is a Luttinger liquid, the $\hat{O}$'s acquire
anomalous
dimensions relative to the non-interacting fixed point.
The successive application of these operators generate
interactions which take a pair of electrons from one set of chains
into the other set and back again. Thus, upon scaling towards low
energies, interactions of the type $\hat{A}_{hh}$ and $\hat{A}_{vv}$
will also be generated. In general, the operator $\hat{A}$ has to be
considered a $2 \times 2$ matrix. It can be shown \cite{us} that
the $\hat{O}$ associated with processes {\it ii)} is less relevant
than the operator which describes processes of type {\it i)}.
Note that, as our scheme implies a summation to all orders of intra-chain
effects, the propagators shown in Fig.\ref{diagram} acquire anomalous
dimensions \cite{boso}.
The scaling equation for the matrix $\hat{A}_{ij}$ has the {\it Riccati form}:
\begin{eqnarray}
\partial_{l} \hat{A} = - \tilde{c}_1(g,l) \tilde{t}^2 \sigma_x
+ 2 \tilde{c}_2(g) \hat{A} - \tilde{c}_3(g,l) \hat{A}^2
\label{scaling}
\end{eqnarray}
where $\sigma_x$ is a Pauli matrix.
$\tilde{c}_1(g,l) = c_1 e^{4 (g-1) l}$ represents the generation of
the Cooper pair tunneling starting from the tunneling of
individual electrons. In a conventional junction between superconductors,
this is the Josephson coupling ($c_1>0$ unless the tunneling takes place
through
a magnetic region \cite{SK}). The second term is
the scaling dimension of the operators with $\tilde{c}_2>0$ ($\tilde{c}_2<0$)
for attractive (repulsive) intra-chain interactions.
$\tilde{c}_3(g,l) = e^{2(g-1)l}$ is associated with
the two-dimensional pairing on the Fermi surface in Fig.\ref{fermi}(a)
\cite{shankar}.
Observe that due to symmetry $\hat{A}=A I + B \sigma_x$
where $A$ and $B$ are real numbers and $I$ is the identity matrix.
It is clear from equation (\ref{scaling}) that it is useful to work
with the eigenvalues of $\hat{A}$:
\begin{eqnarray}
\tilde{\lambda}_{\pm} = A \pm B \, .
\label{eigenvalue}
\end{eqnarray}
After a redefinition of the variables ($\lambda_{\pm}= e^{2 (g-1)l}
\tilde{\lambda}_{\pm}$, $\tau = e^{3 (g-1)l} \tilde{t}$) the RG equations read:
\begin{eqnarray}
\partial_{l} \lambda_+ &=& -c_1 \tau^2 + 2 c_2(g)  \lambda_+ - \lambda_+^2
\nonumber
\\
\partial_{l} \lambda_- &=& c_1 \tau^2 + 2 c_2(g)  \lambda_- - \lambda_-^2
\nonumber
\\
\partial_{l} \tau &=& c_4(g) \tau
\label{rgflow}
\end{eqnarray}
where $c_2(g) = g-1+\tilde{c}_2$ and $c_4(g)=3(g-1)+\tilde{c}_4(g)$.
The RG flow has four fixed points: $(0,0,0)$, $(2 c_2,0,0)$, $(0,2 c_2,0)$ and
$(2 c_2,2 c_2,0)$. The $(0,0,0)$ is the non-interacting fixed point;
$(2 c_2,2 c_2,0)$ is the non-interacting fixed point of isolated
chains but with superconducting fluctuations. The physics of the other two
fixed points
is more complex. To gain more insight we study the RG flow when
$\tau=0$.

\begin{figure}
\epsfysize5 cm
\hspace{0cm}
\epsfbox{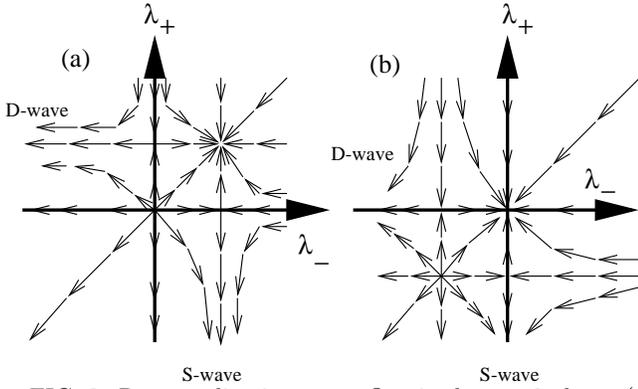}
\caption{Renormalization group flow in the $\tau=0$ plane:
(a) attractive interactions; (b) repulsive interactions.}
\label{flow}
\end{figure}

For attractive interactions the flow is shown in Fig.\ref{flow}(a). Observe
that
$(2 c_2,2 c_2,0)$ is stable whereas the non-interacting fixed point is
unstable;
the other two fixed points are also unstable.
$(2 c_2,0,0)$ is unstable in the negative $\lambda_+$ direction and from
the diagram of Fig.\ref{diagram}(b) we see that the
instabilities is of the superconducting type.
One can understand
the strong coupling fixed point in a very simple way: observe that $\lambda_-$
is constant and $\lambda_+<<-1$. From (\ref{eigenvalue}) we have $A<B<<-1$.
It implies a superconducting instability in which the
sign of the order parameter is the same along horizontal and vertical stripes;
this is equivalent to {\it s-wave} pairing.
$(0,2 c_2,0)$ is unstable in the negative $\lambda_-$ direction while
$\lambda_+$ is constant which implies $A<<-1$ and $B>>1$.
The coefficients have
opposite sign which means that the sign of the order parameter in the
vertical stripes
is opposite to the sign in the horizontal stripes leading to what is
called a {\it $\pi$-junction} \cite{SK}. Since the stripes are oriented
along the crystallographic directions this is equivalent to
{\it d-wave} pairing. For repulsive interactions the s-wave and d-wave pairing
persist
and the flow is shown on Fig.\ref{flow}(b).
If $\tau \neq 0$
the hopping is relevant (except for strongly attractive interactions).
If $c_1>0$ the system flows towards
s-wave and
if $c_1<0$ the flow is reversed towards d-wave superconductivity.
One also expects that in the presence of repulsive interactions
the mesh should be {\it insulating} because of the intrinsic disorder in the
CuO$_2$ planes
\cite{bob}.
A mesh like the one proposed here is very sensitive to disorder and
impurities,
such as Zn. Note that Zn reduces drastically the superfluid density
\cite{balatsky}.
The attraction within each stripe is provided by the
surrounding antiferromagnet \cite{steve,gui}. The model proposed
here should also be useful in describing
artificial Josephson arrays \cite{tinkham}.

Since the flow is divergent at some finite scale $l^*$ we
estimate the gap from the scaling equations
by studying the flow close to the singularity.
At this point the tunneling amplitude is constant and given by
$\tau_*=\tau(l^*)$.
If $\tau_*^2>c_2/c_1$ the gap has a BCS form,
\begin{eqnarray}
\Delta=\omega_p
\exp\left\{-\frac{\frac{\pi}{2}
- \arctan\left(\frac{|\lambda_0-c_2|}{
\sqrt{c_1 \tau^2_*-c_2^2}}\right)}{\sqrt{c_1 \tau^2_*-c_2^2}}\right\}
\label{gapbcs}
\end{eqnarray}
where $\lambda_0$ is the coupling
constant and $\omega_p$ is the characteristic pairing energy.
In this limit one expects usual BCS behavior with well defined
quasiparticles in the superconducting state.
Eq.~(\ref{gapbcs}) reduces to the usual BSC gap ($\Delta_{BCS} = \omega_p
e^{-1/|\lambda_0|)}$) in the non-interacting limit ($c_{1,2} \to 0$).
In the opposite limit,
$\tau_*^2<c_2/c_1$, Luttinger liquid behavior is obtained with a
{\it non-BCS gap} given by
\begin{eqnarray}
\Delta = \omega_p
\left[\frac{\lambda_0-\sqrt{c_2^2-c_1 \tau^2_*}-
c_2}{\lambda_0+\sqrt{c_2^2-c_1 \tau^2_*}-c_2}\right]^{
-\left(2 \sqrt{c_2^2-c_1 \tau^2_*}\right)^{-1}}
\nonumber
\end{eqnarray}
which, when $c_1 \to 0$, has the interesting form
$\Delta_L = \omega_p (1+c_2/|\lambda_0|)^{-1/c_2}$ which raises much
faster with the coupling constant than the usual BCS result.

Let us consider the tunneling of Cooper pairs
between transverse stripes {\it A} and {\it B} through the antiferromagnetic
regions.
Since the pair is in a singlet state the tunneling process occurs in
such a way that a given spin order the sign of the wavefunction is
reversed \cite{SK}. This is the so-called a $\pi$-junction and
implies that $c_1<0$ in the RG flow.
The argument is valid if the tunneling occurs
through a region smaller than the
magnetic correlation length, $\xi_M$ \cite{zeits}. A similar
mechanism for d-wave pairing involving tunneling
through a finite antiferromagnetic region
was found by Bonesteel \cite{bone}.
Assuming that $c_1 \leq 0$
the tunneling through the antiferromagnet only takes place
if $\xi_M(T^*) \approx a$.
Thus, there is a crossover temperature $T^*$ at above which $c_1=0$
\cite{gabe}.
$T^*$ can be roughly estimated assuming that the antiferromagnet is in
the quantum critical regime \cite{zeits} where $T^* \approx c/a$
($c$ is the spin wave velocity).

In conclusion, besides the shape of the Fermi surface,
the existence of ``shadow" bands,
the co-existence of Fermi liquid and non-Fermi liquid behavior,
the possibility of non-BCS and BCS
behavior and s- and/or d- wave pairing
we have three different temperature scales in this problem,
namely, $T^*$, $\delta$ and $\Delta$. Consider first the case
such that $T^* > \delta > \Delta$. If $T^* > T > \delta$
the pairing takes place but there is no coherence between
the stripes. There are pre-formed pairs and
a gap to spin excitations due to the formation of singlets.
This effect is observed experimentally in infrared absorption
and nuclear magnetic resonance (NMR) \cite{infra}.
If $\delta>T$ tunneling takes place
and a gap opens at the $(\pm \pi/a,0)$, $(0,\pm \pi/a)$
regions of the Fermi surface but no phase coherence is
observed. This case has similarities to a recent
model proposed by Geshkenbein {\it et al.}
for the physics of the Fermi surface with pre-formed pairs
\cite{gesh}.
In the other scenario, $\delta > T^* >\Delta$, the physics
is quite different. For $\delta>T>T^*$ a Luttinger liquid
is formed close to $(\pm \pi/a,0)$, $(0,\pm \pi/a)$ but no Cooper
pairs are found.
When $T^*>T$ the Luttinger liquid regions of the
Fermi surface become gapped due to pair formation.
At lower temperatures superconductivity is obtained. While the
first case is closer to a conventional superconducting instability
the last one shows direct transition from a Luttinger liquid to
a superconductor \cite{anderson}.
This picture agrees with the available data in the underdoped cuprates
\cite{shen,randeria}
and provides a microscopic origin for the boson modes proposed by Shen and
Schrieffer
\cite{schrieffer} in terms of collective modes of the Luttinger liquid.
Observe that in the underdoped compounds we predict weak dependence of
$k_F$ on doping \cite{zeits}.

Finally, we can 
predict a new effect as we change the
ratio $r \approx T^*/\delta \approx (c/v_F) (\ell/a)$. Close to
the metal-insulator transition one expects
$r>>1$ because $c$ can be large and $\ell/a \approx 1/x >>1$.
In this case the first scenario applies and as the system is
cooled down a spin gap must be observed before
a gap opens at $(\pm \pi/a,0)$, $(0,\pm \pi/a)$ in the Fermi surface.
At larger doping (but smaller than $x^*$), that is, when $r<<1$, one expects
that
no gap will appear in the Fermi surface (because of Luttinger behavior)
and at lower temperatures a spin gap appears
before the system superconducts.

We thank
G.~Aeppli, L.~Balents, D.~Baeriswyl, G.~Castilla, A.~Chubukov,
M.~P.~A.~Fisher, E.~Fradkin, N.~Hasselmann, S.~Kivelson,
C.~Nayak and S. ~White
for useful discussions, and
the Benasque Center for Physics, for
its hospitality. A.~H.~C.~N. is an Alfred P.~Sloan fellow. F. G. is thankful
to CICyT (Spain) for financial support (grant PB96-0875).

{\bf Note}: Since completing this work, we became aware of numerical
work on stripes in the t-J model\cite{WS98}
which seems to be consistent with our results.

\end{document}